# An investigation of a new metamaterial composite structures

Tao Jiang

A frequency selective device, which is composed of a kind of composite structure, is designed to control the propagation of free space electromagnetic waves in the X band of the microwave spectrum. The composite structure mainly consists of a simple frequency selective surface (FSS) structure and the designed plasma arrays. Firstly, the electromagnetic property of the FSS structure is tested, and results indicating that the FSS structure with coupling reversed phase patternings has a significant absorptivity, with a bit blue shift of the peak resonance in comparison with the single layer FSS cases. Then the designed frequency selective device is still investigated in X band frequency. Results demonstrate that this device is easy to control the propagation of microwave, and the bandgap frequencies can be shifted through varying the FSS structure or the plasma arrays, and that the properties of this structure is highly dependent on frequency.

Frequency selective surface (FSS) is a kind of metamaterial that usually composed of metallic elements and spaced by dielectric substrate, sometimes, these basic elements are arranged in a single layer or multilayers. The original idea is that, for the FSS devices the nearly unity property comes from the impedance matching between the surface Z(w) and the free space $Z_0$.[1] The above impedance matching theory can be realized by tuning the effective electric permittivity $\varepsilon(\omega)$ and magnetic permeability $\mu(\omega)$ independently, as the frequency dependence impedance was determined by equation $z(\omega) = \sqrt{\mu(\omega) / \varepsilon(\omega)}$. To date, numerous of researchers have devoted themselves into this relevant area, and their research findings are significant for the investigation of novel physical phenomena while these findings will also hold great promise for future researches and applications.[2-4]

In many literatures, in order to get proper resonant frequency and obtain the frequency selective controlled device that various approaches have been tried.[5, 6] Based on the theory that the resonance of the electric field and magnetic field can active electrons, and the electric resonance and magnetic resonance stem from the metal structure. We can find that researches are mainly focused on the configuration of metallic patterns' shape and their arrangement on substances.[7-10] The proper structural design for the resonators is a general approach to effectively manipulate the transmission of electromagnetic wave and to get the expecting results.

Though people have given much attention to the structure design of resonators, the property improvement is restricted under single aspect efforts, such as just add the number of wires or minimize the thickness of metamaterials.[1, 11] It is believed that not only the appropriate structural design but also the proper combination of structures can realize the controlling of the effective electromagnetic parameters and result in the predesigned electromagnetic response. In recent years, some researchers investigated the composite structures in the absorption for the EM wave and they got the expected results.[12, 13]

Based on the composite structure idea, we design a frequency selective device that consists of frequency selective surface (FSS) and plasma arrays. Plasma arrays can also react with the EM wave and it can control the propagation of EM wave. It's a fact that the plasma electrons collide with each other and the heat generation is inevitable, so the density of the plasma tubes is limited.

Electromagnetic wave traverses into the FSS, the resonance of the electric field and the magnetic field could activate the electrons, induced current also generates on the metal surface of the FSS. When the EM wave interacts with electrons, it scatters and attenuates resulting in the change of its transmission properties. Researches show that the number of concentric closed patterns on one layer, as well as the number of unit layers have a big impact on the electromagnetic properties and the transmission property of the EM wave. For instance, Li Huang et.al designed the I-shaped resonator, and the optimized three I-shaped resonators exhibited the broadest bandwidth compared with the one I-shaped resonator and even the two I-shaped resonators.[14] N.I.Landy et.al investigated the effect of adding multiple metamaterial layers, their results showed that the absorbance rose sharply with additional layers and was asymptotic to unity. [1] And that, two layers of the absorber could achieve a value of as high as 99.9972%.

One can find that the previous studies mainly focused on periodic units distributed on dielectric base, namely the mechanical reconfiguration of the lattices or the unit cells on the substrates.[5] In this paper, we firstly try to investigate the designed FSS structure which consisting of continuous patterning units. Then we explore the behavior of the EM wave traversing into the composite structure that composed of the designed FSS structure and plasma arrays.

Based on the former researchers, we have two layers with continuous patterning units in the FSS structure, and they are placed in reverse phase paralleling each other. The schematic diagram of this FSS structure is displayed in figure. 1(a), as can be seen that each single layer consists of two components, namely the FR4 sheet (in green color) lies at bottom as substrate and the copper patterning wire (in yellow color) which is printed on the top of the substrate. In this experiment, the coupling patterning pairs with the ground plane, it will induce current in the section of resonators when interacting with the EM wave. In this process, the electric and magnetic response are determined by the geometry of the patterning and the distance between these two metallic components.[15] The geometry of the patternings and their parameters are determined after the optimization and simulation of the FSS structure by the commercial simulation software CST.

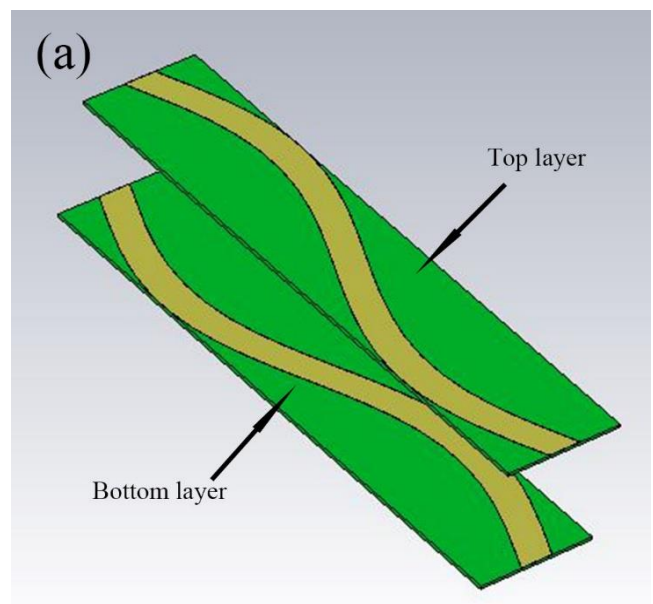

(a)

Top layer

Bottom layer

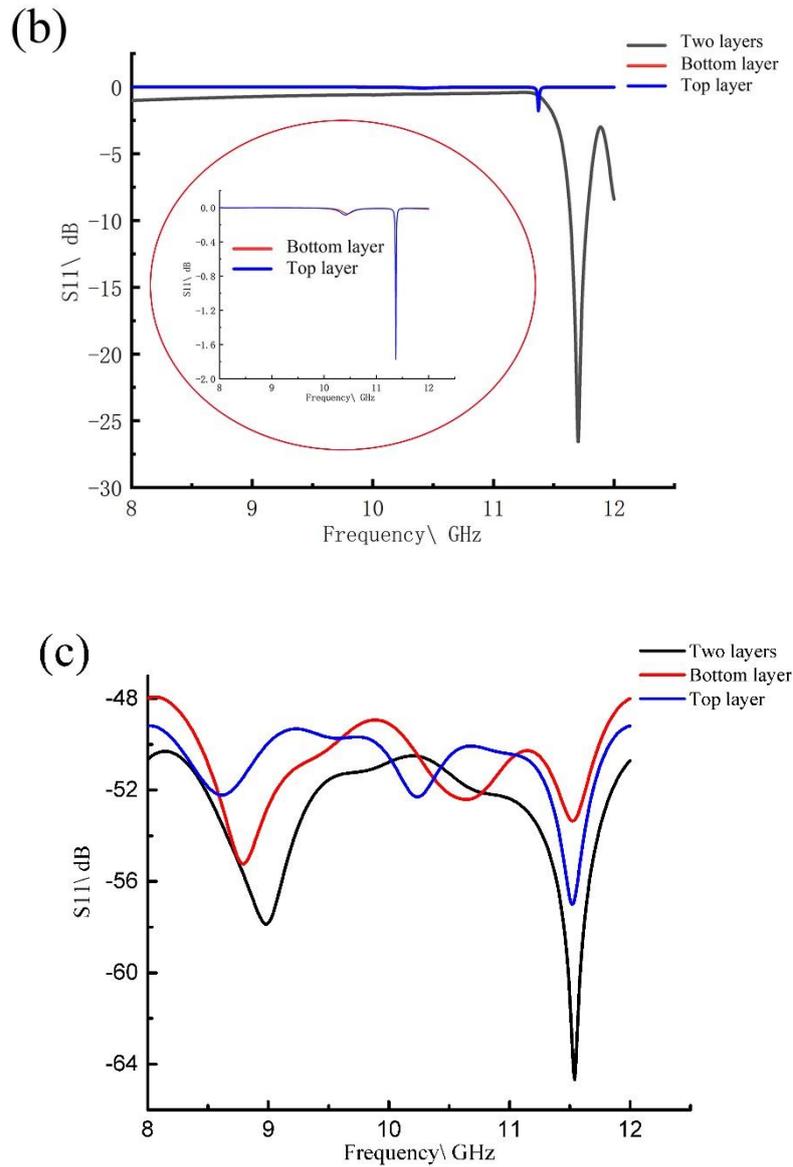

Figure. 1. The frequency selective surface structure: (a) schematic diagram of the patterning units and their relative position in resonator, (b) display of the analog results and, (c) the tested results for the reaction between microwave and the FSS structure.

In the first place, we perform some numerical simulations, meanwhile do some corresponding tests to verify the early hypothesizes. Figure. 1(b), (c) illustrates the results of simulations and experiments for the reaction between the FSS structure and the EM wave in X band frequency. In figure. 1(b), the present dominant coordinate system shows that when only one layer (the top layer or the bottom layer) existed, the impact between the resonator and the EM wave is quite weak, leading to the low attenuation of the EM wave but a high reflectivity. Whereas, when the two layers exist simultaneously the EM wave interacts with resonators strongly and it attenuates too much, eventually, the reflectivity decreased sharply. Details of impact of the existence of single layer is presented in red circle of inset in figure. 1(b), it tells that the two graphs are almost the same at X frequency band. In other words, in the case of only one layer existing the coupling direction makes no difference on the electromagnetic property of the EM wave. The corresponding testing results are supplied in figure.1 (c). There is a large resonance peak at about 11.5GHz, which is consistent with the analog results considering the existing of the errors.[16] We can find that in the testing graphs, a small resonance peaks occur at the low frequency and some fluctuations vibrate in the midband. This is mainly due to the errors of the realistic testing samples in the experiment, such as the surface bending and the boundary sharpening of the FSS, which may lead to the multiple reflection of the EM wave and the current density effect. All these results suggest that in this work the predesigned coupling effect between the top layer and the bottom layer is effective.

In the following parts we will investigate the behavior between the EM wave and the composite structure, and the designed FSS structure which has been mentioned above is one component of the composited structure.

The designed composite structure is displayed in figure. 2, schematically. The geometric dimension, physical properties, the number of patterning unit layers and even their relative position of the FSS structure are the same as discussed in the previous section. In figure.2, the blue cylinders in the FSS structure are commercial fluorescent lamps of T4 type, noting that these fluorescent lamps are named T4s in the following statements. In this experiment, an AC ballast, in this way, discharges one T4 so each T4 can be driven and controlled individually. The distribution of these T4s can be clearly seen in section view part of figure. 2. For the Top layer_comp layer, there are three T4s and each T4 is placed just between the two coupling patterning units of the FSS structure. In the Bottom layer_comp layer the number of T4 is five, and they follow the top and bottom alignment distribution. Discharge these T4 tubes in a certain array can form a kind of structure which named photonic crystal, when the EM wave propagates into these structures they will interact with each other.[17]

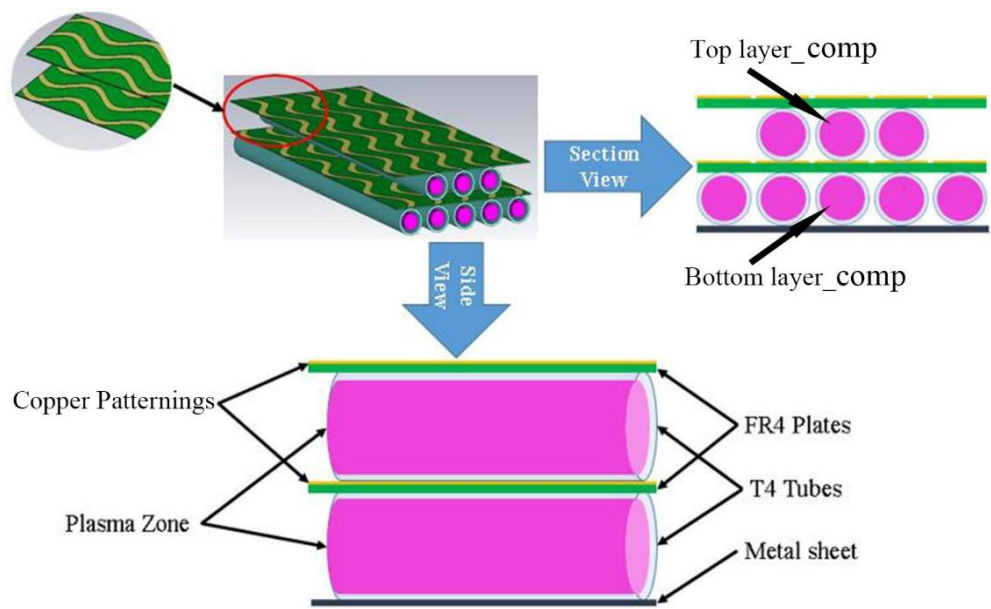

Figure. 2. Schematic diagram of the metamaterial composite structure and the distribution of its components in section view and in side view.

We perform this experiment in the self-designed anechoic chamber in the key laboratory, and figure.3 shows the test-site of the experiment. From the site picture, we can see that our sample size is 80mm*120mm, and the distance from the antennas to the testing target designed small is about 870mm. Absorbing sponges are orderly placed on the floor and well organized around the testing samples zone, they can restrain the interference from the electromagnetic reflection inside the chamber, and shield the electromagnetic interference from outside simultaneously. In this experiment, a pair of microwave broadband horns are hanging just above the testing board serving as the source and the detector. The type of the antenna is LB-10180-NF and its working frequency range is 1.0-18.0GHz. An Agilent 8720ES S-Parameter Network Analyzer is connected to the antenna to measure the transmission coefficient.

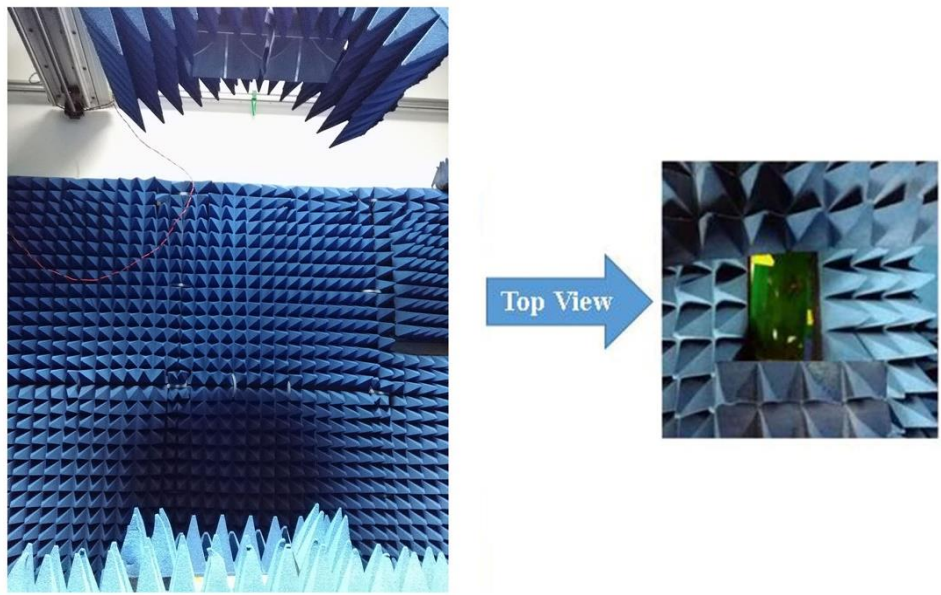

Figure. 3. The testing site of the reflectivity of the interaction between metamaterial composite structure and the EM wave.

As have been mentioned above that each T4 can be controlled individually. In the experiment, these eight T4s are divided into two groups, namely three T4s in the Top layer_comp layer are one group and the other five are another group. In the following experiments, these T4s are charged or discharged in groups. As the fact is that the size of test target and the frequency of microwave have a great influence on the accuracy of the testing results.[18] To achieve an effective test, we choose and set the testing frequency range of the antenna at higher frequencies between 8.0GHz to 12.0GHz. Firstly, we discharge three T4s in the Top layer_comp layer as one situation and mark as Top layer_comp, and then trigger the antenna that serves as microwave emit source and detector device. Then after the test data are stored, the current of this group is cut off and another group discharges as the second situation marking as Bottom layer_comp. Similarly but not identically, when the second data has saved, the former group will discharge again but keep the just discharging group on this time as the third situation and marks as Two layers_comp. It is the truth that not all T4s could reach the steady discharge simultaneously because of the individual differences among T4s. Therefore, T4s discharged about five seconds before triggering the test in order to obtain the same plasma in every T4 of each situation.

Figure. 4(a), (b) show the experimental reflectivity spectra S11 for three cases of the composited structure and only the plasma arrays existed in 8.0-12.0GHz frequencies, respectively. These experiments were performed at incidence angle of 90° with the microwave electric field paralleling to metallic patterns. There are many factors contributing to the testing errors, among them the calibrating of network

analyzer system can greatly improve the measurement accuracy.[18] In figure. 4(a), lines of Bottom layer_comp and Top layer_comp correspond to the two situations of only one group of T4s are discharged in this experiment. When only the Bottom layer_comp group discharged, blue-shift occurs in this condition. What's more, the resonant point appears in the midband, but the other resonant peaks disappear in low frequency and high frequency in compared with the testing results, which is illustrated in figure. 1(c). For the group of Top layer_comp, the midband resonant peak is gone and in the high frequency red shift appears. The same reflectivity for the EM wave as group Bottom layer_comp is that the value is very low.

The fact is that when the EM wave propagates in the vacuum, it follows the Maxwell equation that given as

$$\nabla * E = -\mu_0 \frac{\partial H}{\partial t} \tag{1}$$

$$\nabla * H = \varepsilon_0 \frac{\partial E}{\partial t} + J \tag{2}$$

and the constitutive relations are given as

$$D = \varepsilon \varepsilon_0 E \tag{3}$$

$$B = \mu \mu_0 H \tag{4}$$

Where E and H are the electric and magnetic field of the EM wave, $\mu_0$ and $\varepsilon_0$ are the permeability and permittivity in vacuum, t is the time, $J$ is the external current density in Maxwell equation. Additionally, because the patterning units of the FSS structure in the composite structure couples in reverse phase, there is the induced current on the metasurface of patternings, which found distributing mainly in the edge and the corner.[19-21] In this experiment, the $J$ is the induced current distributed on the metasurface of the patternings. Coupled with the presence of plasma arrays, which has the variable value of permittivity, the electromagnetic FSS structure form such a device. As a response against the electric and magnetic field of the propagating wave, such a microstructure induces anomalous charges and currents in electric properties. Induced electric field is also formed by the induced current in the composite structure. When the EM wave traverses into these zone, it reacts with electrons and the electric field. Then the interaction between microwave and the device is enhanced, and the source of radiation attenuates too much which leads to a low reflectivity.

According to Paschen's law, the higher density of plasmas leads to the electron plasma frequency coming into the frequency range of millimeter waves. What's more, the frequency of the radiation source has significant roles in the resonance,[22] the testing in the last part will also prove it. Electron plasmas react with the microwave scattering and interference, and when the dielectric constant of the composite structure makes its impedance match the free space and the resonance occurs. The detailed mechanism explain for this phenomena is required for further investigation.

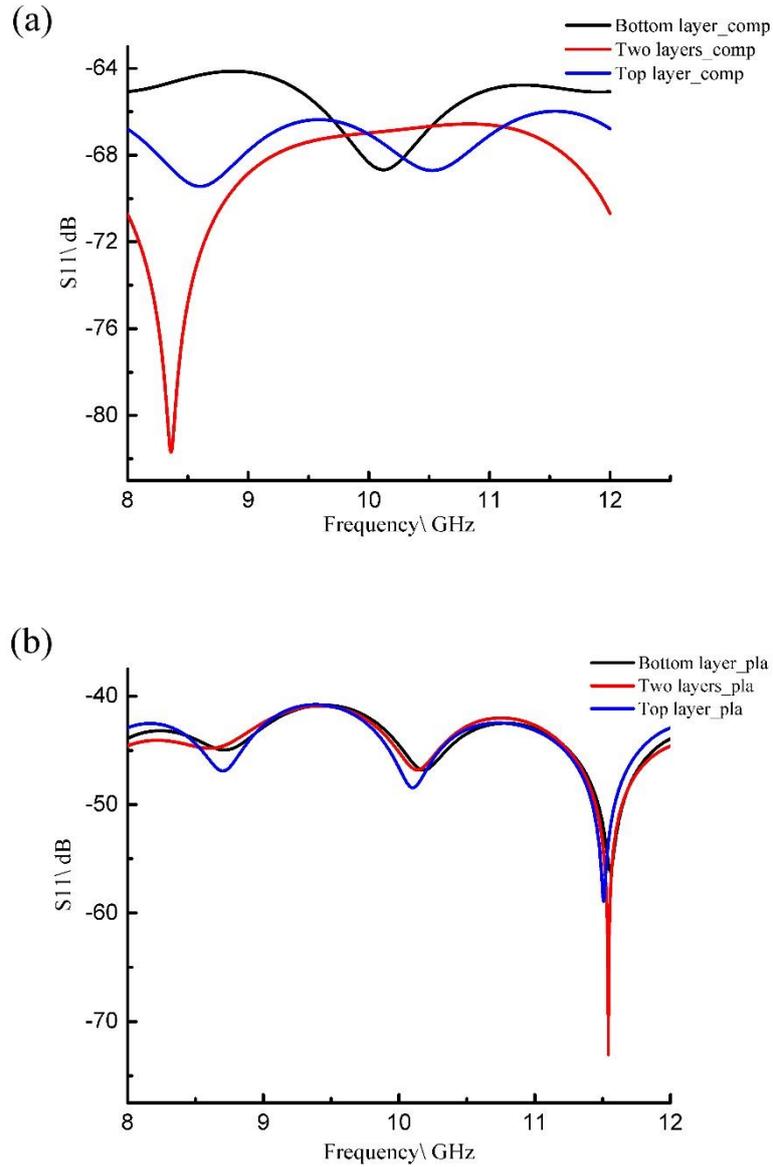

Figure. 4. Reflectivity of impact between, (a) metamaterial composite structure, (b) only the plasma arrays existing, and the EM wave at 8-12GHz.

Placed in the end, however, the Two layers_comp is the most significant changes among the three situations displayed in figure. 4(a). The resonance appears in low frequency at about 8.4GHz, and there is a significant red shift in comparison with the case without the plasma array existed which is shown in figure. 1(b). This is the frequency selective property, which can be used in switch and other applications like filters. This is an inherent typical characteristic of the plasma photonic crystal, one feature that depending on the electron elastic collision. Namely, when the EM wave propagates into plasma zone, it collides with electrons and then scatters and attenuates. Generally, the Drude Model is used to describe the electric constant of the plasmas,

$$\varepsilon_p(\omega) = 1 - \frac{\omega_p^2}{\omega^2 - i\gamma\omega} \qquad (5)$$

$$\omega_p = \sqrt{\frac{n_e e^2}{m_e \varepsilon_0}} \qquad (6)$$

in which $\omega_p$ is the frequency of plasma, $n_e$ is the electron density, and the dielectric constant is a function of frequency and the electron density.[23, 24] When electron plasma interacts with the EM wave the electron temperature rises and results in the changing of the number of electron. Once the electron density gets a new value again, the dielectric constant of the plasma realizes the impedance matching to the free space, and the resonance happens at a new band stop frequency.

As far as frequency dependence is concerned, the EM wave attenuation enhanced with the increase of the density of electron, in this test it presents a large attenuation rate and reflects a high ratio of dielectric constant in plasmas and the free space, eventually leading to a low resonant frequency.[25] In terms of the effect of the FSS structure on the composited structure. When microwave traverses into the composite structures, the induced current in the upper patterning unit layer and in the below layer will be parallel to each other but in opposite direction. This would lead to the magnetic coupling and the reaction with the EM wave, finally, the resonance occurs at the low frequency as the source of the radiation has been weakened a lot by plasmas in the meantime.[21]

Figure. 4(b) illustrates the testing results of the T4s arrays interact with the EM wave in 8-12GHz frequencies. It tells that the plasmas have a stopband for certain microwave, and the strongest resonance occurs when the two layers of T4s are discharged simultaneously. Several coincident fluctuations in the lower frequencies stem from the multiple reflection of microwave by the lamp surfaces. Comparing with graphs in figure. 4(a) and figure. 1(c), we can find that the composited structure, which is composed of the designed FSS and plasma arrays, indeed has a special electromagnetic property that is different from the single composition.

In the end, the behavior of this composite structure reacting with microwave in the lower and higher frequencies is investigated. In these cases, experiment conditions and the experimental process are just as the same as the experiments in 8-12GHz frequency band.

Figure. 5 displays the reflectivity of impact between the composite structure and the other frequency bands of the lower and the higher frequencies in comparison with the above used microwave in X band. Comparing with the results in figure. 4(a), the reflectivity of impact between the composite structure and the EM wave is highly depends on frequency. For the Bottom layer_comp layer, there are two resonant points distributing at the lower band and the higher band in frequencies of 6-8GHz and 12-18GHz respectively. However, the case is on the opposite when it refers to the Top layer_comp layer, as there is only one resonance in 6-8GHz and 12-18GHz. For the situation Two layers_comp, there are three resonant points appearing in each of these three frequency bands respectively. The absorptivity for cases of Bottom layer_comp and Top layer_comp the highest values appear at the situation of 12-18GHz, but for the Two layers_comp case the absorptivity rises as the spectrum increases.

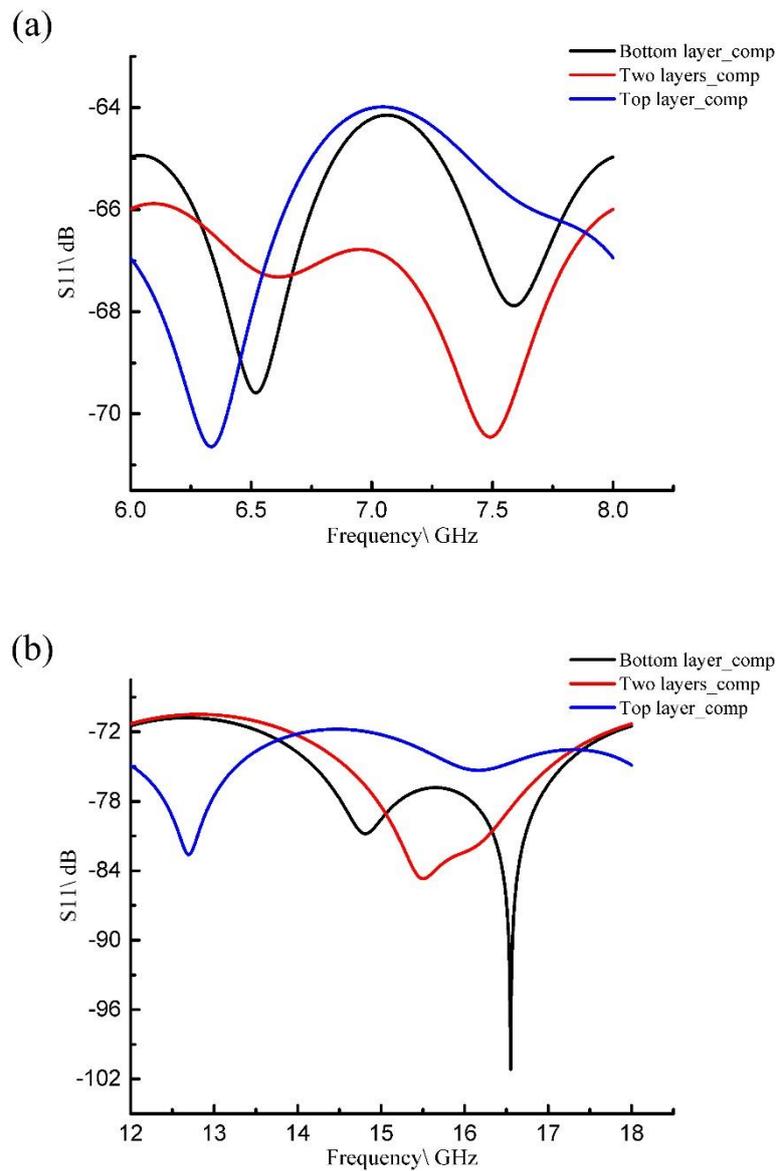

Figure. 5.　Reflectivity of impact between metamaterial composite structure

and the EM wave (a) at 6-8GHz, (b) at 12-18GHz.

4 Conclusions

   In conclusion, we have designed and fabricated a composite structure device consisting of the designed FSS structure and plasma arrays, and for the first time, we have characterized its electromagnetic property as a frequency selective device. Testing results show that, the analog results have a significant guidance to the experiment in designing double layer structure of the FSS interlayers, and that this structure has an effective absorptivity to microwave in X band frequency. For the designed composite structure device, in the Two layers_comp situation, namely two layers of compound fluorescent lamps are discharged, it exhibits a special electromagnetic property comparing with only the designed FSS structure or the plasmas existed, and with a large red-shift in the X band frequency. Furthermore, this case possesses a higher absorbability than the other two cases. The later tests also illustrate that the designed composite structure device has a high dependence on frequency, and this device has a great potential applications in switches and electromagnetic filters. Future works will focus on the effect of discharging certain array of the plasma tubes on electromagnetic property of the composite structure. The characterization and design of this composited structure will provide a new method for researchers in investigating frequency selective devices.